\documentclass[a4paper]{article}

\usepackage{INTERSPEECH2022}
\usepackage{algorithm}
\usepackage{algorithmic}
\usepackage{hyperref}
\usepackage{caption}
\title{Improving Streaming End-to-End ASR on Transformer-based Causal Models with Encoder States Revision Strategies}
\name{Zehan Li$^{1,2}$,Haoran Miao$^{1,2}$,Keqi Deng$^{1,2}$,Gaofeng Cheng$^{1}$,Sanli Tian$^{1,2}$,Ta Li$^{1,2}$,Yonghong Yan$^{1,2,\dagger}$\thanks{\(\dagger\) Corresponding author}\thanks{This work is partially supported by the Goal-Oriented Project Independently Deployed by Institute of Acoustics, Chinese Academy of Sciences(MBDX202106)}}

\address{
  $^1$Key Laboratory of Speech Acoustics and Content Understanding, Institute of Acoustics, CAS\\ 
  $^2$University of Chinese Academy of Sciences, China}
\email{\{lizehan,miaohaoran,dengkeqi,chenggaofeng,tiansanli,lita,yanyonghong\}@hccl.ioa.ac.cn}

\begin{document}

\maketitle

\begin{abstract}
%   Balancing performance and latency of streaming automatic speech recognition (ASR) is often a dilemma. Regular methods like look-ahead and chunk-based method usually require information from future frames to advance recognition accuracy, which incurs inevitable latency even if the computation is fast enough. A causal model that computes without any future frames do not have this inevitable latency, but its performance is much worse than regular methods. In this paper, we propose corresponding revision strategies to improve the causal model. Firstly, we introduce a real-time encoder states revision strategy to modify previous states. Encoder forward computation starts once the data inputs and revises the previous encoder states after several frames, which is no need to wait for any right context. Moreover, a CTC spike position alignment decoding algorithm is designed to reduce time costs brought by the proposed revision strategy. Experiments are all conducted on the Librispeech dataset. Fine-tuning on the CTC-based wav2vec2.0 model, our method can best achieve 3.7/9.2 WERs on test-clean/other sets and brings 45\% relative improvement for causal models, which is also competitive with the chunk-based methods and the knowledge distillation methods.  %tackle (23\% improvement) (6\% improvement)
  There is often a trade-off between performance and latency in streaming automatic speech recognition (ASR). Traditional methods such as look-ahead and chunk-based methods, usually require information from future frames to advance recognition accuracy, which incurs inevitable latency even if the computation is fast enough. A causal model that computes without any future frames can avoid this latency, but its performance is significantly worse than traditional methods. In this paper, we propose corresponding revision strategies to improve the causal model. Firstly, we introduce a real-time encoder states revision strategy to modify previous states. Encoder forward computation starts once the data is received and revises the previous encoder states after several frames, which is no need to wait for any right context. Furthermore, a CTC spike position alignment decoding algorithm is designed to reduce time costs brought by the proposed revision strategy. Experiments are all conducted on Librispeech datasets. Fine-tuning on the CTC-based wav2vec2.0 model, our best method can achieve 3.7/9.2 WERs on test-clean/other sets and brings 45\% relative improvement for causal models, which is also competitive with the chunk-based methods and the knowledge distillation methods.  %tackle (23\% improvement) (6\% improvement)

\end{abstract}

\section{Introduction}
  End-to-end models, including connectionist temporal classification (CTC) \cite{graves2006connectionist,miao2015eesen,deng2022improving,kurata2019guiding}, RNN-Transducer (RNN-T) \cite{zhang2020transformer,rao2017exploring,graves2013speech} and attention-based encoder-decoder (AED) \cite{chorowski2015attention,watanabe2017hybrid,9739972,deng2021improving} models, have achieved great success on various ASR tasks in the past few years. Many recent studies have focused on streaming ASR, especially the Transformer-based model has received extensive interest because of its parallel training and promising results. %Streaming ASR requires waiting for as little right context as possible. 
  To realize streaming encoder, methods can be categorized into the look-ahead based \cite{povey2018time,moritz2020streaming,wang2020low} and chunk-based \cite{dong2019self,miao2020transformer,chen2021developing,miao2020online}. The former sets a look-ahead window providing context information for each current block, but the latency will grow as the number of layers increases. The latter can be trained in parallel, but the performance is poor at chunk boundary, and accuracy drops significantly when reducing the chunk size.
  
  For some tasks that require low average character latency, i.e., spoken dialog \cite{selfridge2011stability} and real-time translation \cite{arivazhagan2019monotonic}, most of the above methods need to wait for right context before output each time to achieve better results. This will lead to relatively high latency and a worse user experience when the speech length is much smaller than the right context. A causal model that does not need any right context but computes and decodes at each frame may help solve this problem. However, causal models often underperform due to the lack of future information; thus, they are usually part of cascade models rather than used alone \cite{narayanan2021cascaded}.% However, directly trained causal models often underperform due to a lack of future information, which may be a reason for fewer studies on Transformer-based causal models.
  
  An obvious idea to improve the causal model is correcting its outputs. Block-wise attention with a mask-predict method is proposed in \cite{wang2021streaming}. CTC first predicts the preliminary tokens per block, and then low-confidence tokens are masked and re-predicted by a decoder. However, this work only involves the decoder part without any modification to the encoder, and the mask-predict operation must be conducted after the entire block has been calculated. In \cite{huang2020dynamic}, authors propose to use dynamic latency and revise the state of encoder and decoder so that both streaming and non-streaming models can apply incremental decoding \cite{liu2020low}. This method revises the states that have not been completed for RNN-T models and can not change the previous information of the encoder either. In practical applications, to avoid distracting the user's attention, it is more desirable to ensure a fast response and previous stable outputs simultaneously \cite{shangguan2020analyzing}. The above methods will significantly change the previous outputs when the encoder states are inaccurate, which can not be used to improve causal models.

  In this paper, we explore a new method to improve causal models, called encoder states revision strategy. Specifically, we first let the model calculate causally until the revision interval to revise the previous states and correct the decoding path. By applying revision, later outputs will be more accurate, and there is no need to change the hypothesis after all final outputs. Furthermore, we design a CTC spike position alignment decoding algorithm to reduce the computation costs, which is applicable to any revision methods that change the decoding path. Our experiments are based on the wav2vec2.0 \cite{baevski2020wav2vec} fine-tuning models using CTC and conducted on Librispeech benchmark \cite{panayotov2015librispeech}. Our best method can achieve 3.7/9.2 WERs which brings tremendous improvement for causal models and can be compared with chunk-based and knowledge distillation methods (23\% and 6\% relative reduction).

\section{Proposed methods}

\subsection{Encoder states revision strategy for causal model}
  A causal model computes the output of \textit{t}-th frame for \textit{l}-th layer only depending on previous results from \textit{l-1}-th layer. To save computation costs, it is necessary to cache the historical outputs \(\mathbf{Y}^{l-1}_{t-h}, \mathbf{Y}^{l-1}_{t-h+1}, ..., \mathbf{Y}^{l-1}_{t-1}\) when streaming decoding, and the output is
  \begin{equation}
      \mathbf{Y}^{l}_t = f(\mathbf{Y}^{l-1}_{t-h}, ..., \mathbf{Y}^{l-1}_{t})
      \label{eq1}
  \end{equation}
  where \(h\) is the number of history states, \(f\) is a mapping function, and other calculation (i.e., projection, feed-forward) is omitted. Non-causal models usually introduce information of future frames when computing the output of \textit{t}-th frame, so the output is changed to
  \begin{equation}
      \mathbf{Y}^{l}_t = f(\mathbf{Y}^{l-1}_{t-h}, ..., \mathbf{Y}^{l-1}_{t}, \mathbf{Y}^{l-1}_{t+1}, ..., \mathbf{Y}^{l-1}_{t+c})
      \label{eq2}
  \end{equation}
  where \(c\) is the number of future states participating in the computation. Since the causal model lacks future information \(\mathbf{Y}^{l-1}_{t+1}, ..., \mathbf{Y}^{l-1}_{t+c}\) at every \textit{l}-th layer, it performs much worse than the non-causal model.
\begin{figure}[t] %htbp
\centering
\includegraphics[width=\linewidth]{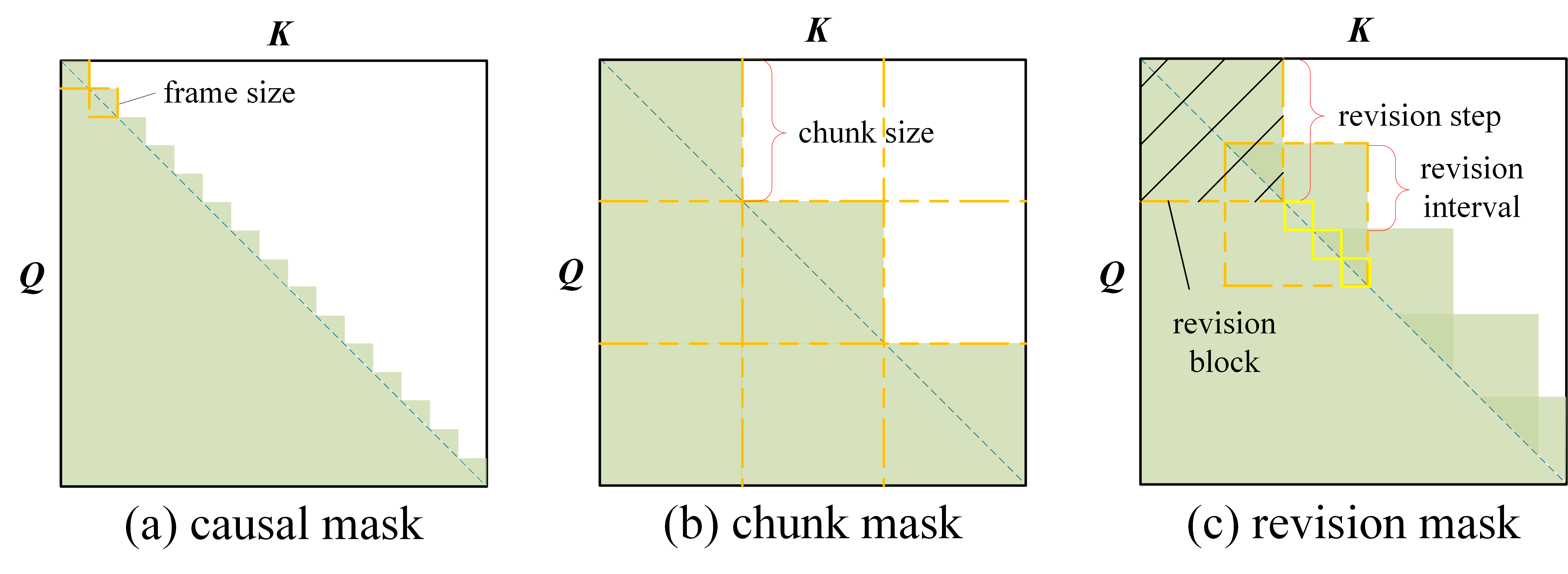}
\caption{Attention mask used in attention mechanism. \(\mathbf{Q}\) and \(\mathbf{K}\) denote the query and key matrix. Green means there is a dependency, while white means there is no dependency.}
\label{fig:attention_mask}
\end{figure}
  
  To improve the causal model, we propose an encoder states revision strategy to add future information. The model outputs causally at first. Once the input frame reaches the index of the revision interval, the model revises encoder states based on previous information of revision steps and outputs a new hypothesis. The model then outputs causally again until the next index of the revision interval and keeps looping. History states before the index of revision step are only involved in the computation and will not be revised. This will add future information for frames before revision interval without too many forward computation costs.
  
  In practical implementation, to ensure the accuracy of real-time outputs, we revise each interval before the first revision step utilizing all the inputs less than the step size. When the index of the first revision step reaches, all previous states are uniformly changed. Then the information on revision step size is used to revise the states every revision interval.
  
  Specifically, we mainly focus on the Transformer encoder in this work. We plot an attention mask in Figure~\ref{fig:attention_mask} for a more intuitive presentation of the revision strategy applied in the attention mechanism. The attention mask limits the range of query-key (\(\mathbf{Q}\)-\(\mathbf{K}\)) dot products for each frame in the attention mechanism. A causal mask used for the causal model is shown in Figure~\ref{fig:attention_mask}(a). \(\mathbf{Q}\) of each frame can only be calculated without any future frames of \(\mathbf{K}\). A chunk mask used for the chunk-based model is shown in Figure~\ref{fig:attention_mask}(b), which allows \(\mathbf{Q}\) to be calculated only with the \(\mathbf{K}\) of the current chunk and the history chunk. Thus, \(c\) in Eq. \ref{eq2} is an integer multiple of the chunk size. Our revision strategy is shown in Figure~\ref{fig:attention_mask}(c). The block at each revision step is called a revision block. Each revision block can be overlapped, which means the frame at the revision interval index will also be revised, and future information will be added in the subsequent revision operation. \(\mathbf{Q}^l_t\in \mathbb{R}^{d_q}\), \(\mathbf{K}^l_t\in \mathbb{R}^{d_k}\), \(\mathbf{V}^l_t\in \mathbb{R}^{d_v}\) denote query, key and value vectors of \textit{l}-th layer at \textit{t}-th frame, \(d_\ast\) denotes dimensions, \({d_q} = {d_k}\). Then the cached states for \textit{l}-th layer are \(\mathbf{K}^l_{t-h}, ..., \mathbf{K}^l_{t-1}\) and \(\mathbf{V}^l_{t-h}, ..., \mathbf{V}^l_{t-1}\). We convert the concatenation operation into a summation symbol, thus in one attention head the output of \textit{l}-th layer at \textit{t}-th frame is 
  \begin{equation}
      \mathbf{Y}^{l}_t = \sum_{j=t-h}^{t}\frac{ e^{\alpha^{l}_{t, j}} }{ \sum_{i=t-h}^{t}e^{\alpha^{l}_{t, i}} }\mathbf{V}^{l}_j
      \label{eq14}
  \end{equation}
  \begin{equation}
      \alpha^{l}_{t, j} = \frac{ {\mathbf{Q}^{l}_t}{{\mathbf{K}^{l}_j}^T} }{\sqrt{d_k}}
      \label{eq15}
  \end{equation}
  
  If the revision step and interval size are \(\sigma\) and \(\nu\), the current index of revision interval is \textit{n}, then the index of states from \(n-\sigma+1\) to \(n-1\) will be revised at \textit{n}-th index of revision interval. The causal output of \textit{l}-th layer after revision is
%   \begin{align}
%   \notag
%       \mathbf{Y}^{l}_m &= \sum_{j=h}^{n-\sigma}\frac{ e^{\alpha^{l}_{m, j}} }{ \sum_{i=h}^{n-\sigma}e^{\alpha^{l}_{m, i}} }\mathbf{V}^{l}_j + \sum_{j=n-\sigma+1}^{n-1}\frac{ e^{\beta^{l}_{m, j}} }{ \sum_{i=n-\sigma+1}^{n-1}e^{\beta^{l}_{m, i}} }\mathcal{V}^{l}_j \\
%       &+ \sum_{j=n}^{m}\frac{ e^{\alpha^{l}_{m, j}} }{ \sum_{i=n}^{m}e^{\alpha^{l}_{m, i}} }\mathbf{V}^{l}_j
%       \label{eq151}
%   \end{align}
  \begin{equation}
      \mathbf{Y}^{l}_m = \sum_{j=t-h}^{n-1}\frac{ e^{\beta^{l}_{m, j}} }{ \sum_{i=t-h}^{n-1}e^{\beta^{l}_{m, i}} }\mathcal{V}^{l}_j + \sum_{j=n}^{m}\frac{ e^{\alpha^{l}_{m, j}} }{ \sum_{i=n}^{m}e^{\alpha^{l}_{m, i}} }\mathbf{V}^{l}_j
      \label{eq151}
  \end{equation}
  where \(m\in[n, n+\nu)\), \(\beta^{l}_{m, j} = \frac{ {\mathbf{Q}^{l}_m}{{\mathcal{K}^{l}_j}^T} }{\sqrt{d_k}}\) and \(\mathcal{K}, \mathcal{V}\) represent all the revised states before \textit{m}-th frame.

\subsection{CTC spike position alignment decoding}
    \begin{figure}[t]
    \centering
    \includegraphics[width=\linewidth]{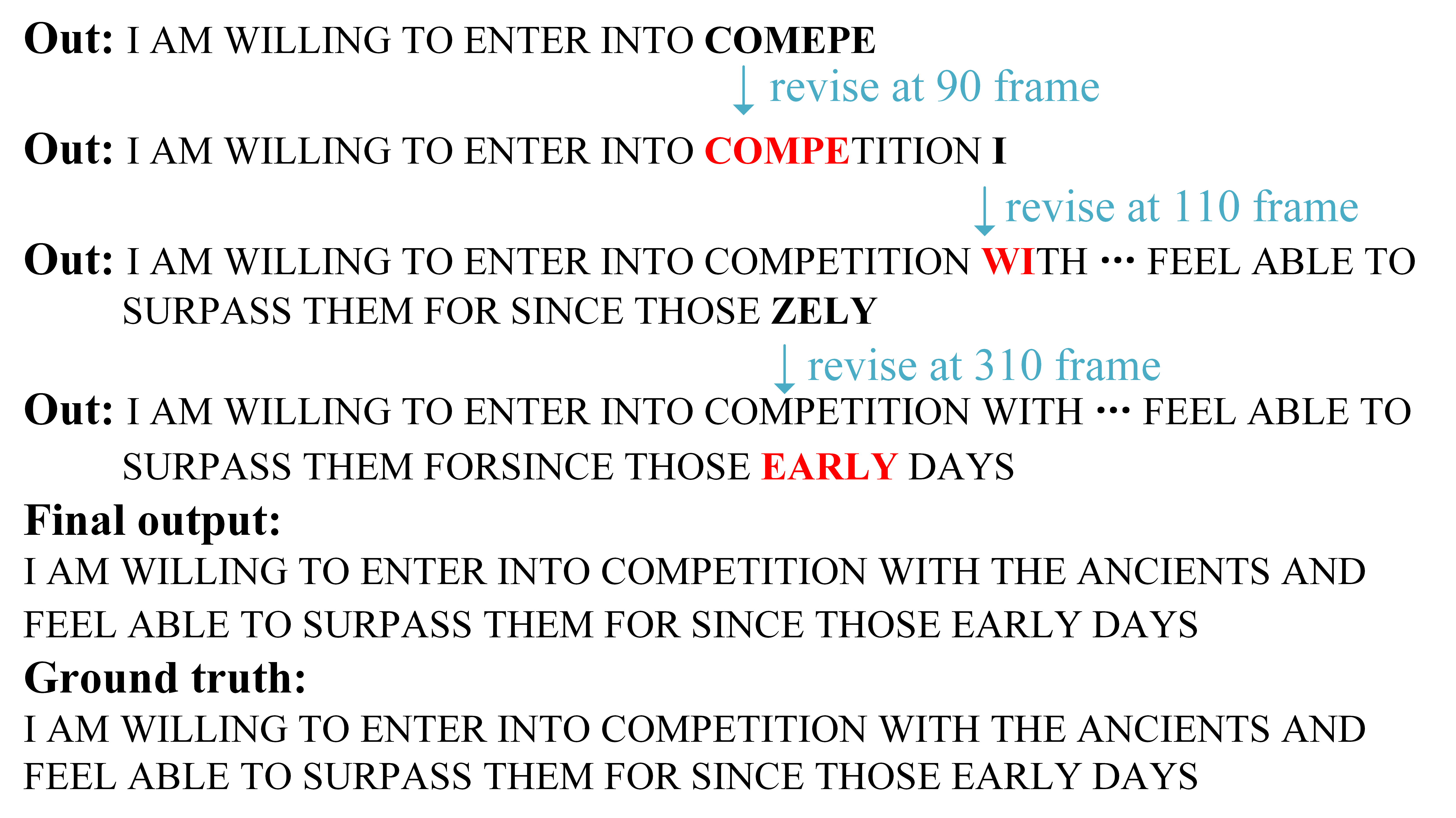}  %0.9\linewidth, height=0.3\linewidth
    \caption{An example of actual outputs. This example is only based on CTC greedy search to exclude the effect of LM, and some revision results are omitted. \textit{COMEPE} changes to \textit{COMPE} at 90 revision frame, and later output is based on the revised states thus can correctly output \textit{COMPETITION} without any further adjustment.}
    \label{fig:output}
    \end{figure}

  The CTC probability may be modified when encoder states change, which affects the decoding path. Therefore, it is necessary to re-decode the output of the revision step. We assume ordinary decoding time is \textit{T}, then when re-decoding, the time costs of the encoder revision strategy will increase
  \begin{equation}
      \Delta T = \lceil \frac{T-\sigma}{\nu} \rceil \sigma + 0.5(\mu + \mu^2)\nu + \eta \sigma
      \label{eq16}
  \end{equation}
  where \(\mu = \lfloor \frac{\sigma}{\nu}\rfloor\), \(\eta \in \{0, 1\} \) and the second part of the formula is revision time costs before the first revision step. \(\eta\)=1 if we apply an additional revision after the final outputs to improve the performance at the end of sentences. This will cause enormous computation time costs if not applying any decoding strategy. %If not applying any decoding strategy, this will cause enormous computation time costs.

  Since \(\sigma\) frames are recomputed, and the length of the causal output is only \(\nu\), most of the positions between causal and revision outputs are likely the same in the beginning. Furthermore, the CTC spike positions in two types of outputs may not be aligned, yet the labels corresponding to their spikes may be identical. In Algorithm~\ref{alg:decode}, we present a suitable \textit{CTC Spike Position Alignment Decoding} algorithm to avoid unnecessary modifications and find the decoding path which is authentic to be changed during frame-by-frame comparison.

  If the current revision frame size is \(\mathcal{T}\), then the revision will change the entire \(\mathcal{T}\)-1 frames of previous outputs, and the current output is at \(\mathcal{T}\)-th frame. We define \(b^1_t, b^2_t\) as the new highest and second-highest probabilities label. Then we use the relative value of \(b^1_t, b^2_t\) to get the dominant label \(\varphi_t\). \(\varphi_t=b^1_t\) if \(\frac{b^1_t - b^2_t}{b^1_t} \geq \theta\) and \(\varphi_t=-1\) if above condition is not met. \(\theta\) is a threshold that varies between 0 and 1. The previous label \(a^1_t, a^2_t\) and the previous dominant label \(\psi_t\) are also obtained similarly. \(\left\langle blk \right\rangle\) represents the blank label in CTC. In lines 3-13, we discuss the case when \(a^1_i = b^1_j\). If they are all the dominant label, or if both are not but \(a^2_i = b^2_j\), this means the outputs are approximately the same, whereas results need to be modified under other conditions. In lines 14-25 is the case where \(a^2_i \neq b^2_j\). When there is a \(\left\langle blk \right\rangle\) dominant label, we need to find the next label which is not \(\left\langle blk \right\rangle\). In line 16, function \textit{FindNext} tries to capture the next same dominant label and return their positions. If there is no dominant label in one part, and the largest label is \(\left\langle blk \right\rangle\), we observe whether the largest label of another part is the same as the second-largest label of this part. We will find the next label if they are the same. If neither of them has a dominant label, this frame is considered to be modified since the largest label is different. The \(\mathcal{T}\)-th frame is a new input, so if \(\tau\) is still zero after the entire loop, which means that all the \(\mathcal{T}\)-1 frames are equivalent, then we only propagate on \(\mathcal{T}\)-th frame. When propagating, we use the phone synchronous decoding (PSD) method proposed in \cite{chen2016phone}, which only expands the \(\left\langle blk \right\rangle\) node when it is the dominant label.
  
\vspace{-0.3cm}
    \begin{algorithm}[h]
    \caption{CTC Spike Position Alignment Decoding}
    \label{alg:decode}
    \begin{algorithmic}[1]
        \REQUIRE{previous top two labels $\{(a^1_1, a^2_1), ..., (a^1_{\mathcal{T}-1}, a^2_{\mathcal{T}-1})\}$, new top two labels $\{(b^1_1, b^2_1), ..., (b^1_{\mathcal{T}-1}, b^2_{\mathcal{T}-1})\}$, previous dominant label set $\Psi = \{\psi_1, ..., \psi_{\mathcal{T}-1}\}$, new dominant label set $\Phi = \{\varphi_1, ..., \varphi_{\mathcal{T}-1}\}$}
        \ENSURE{beginning index $\tau$ for re-decoding}
        \STATE $i \leftarrow 1$, $j \leftarrow 1$, $\tau \leftarrow 0$
        \WHILE{$j < \mathcal{T}$}
            \IF {$ a^1_i = b^1_j $}
                \IF {$\psi_i, \varphi_j \neq -1$}
                    \STATE $i \leftarrow i+1$, $j \leftarrow j+1$
                    \STATE $continue$
                \ELSIF{$\psi_i, \varphi_j = -1$ and $a^2_i = b^2_j$}
                    \STATE $i \leftarrow i+1$, $j \leftarrow j+1$
                    \STATE $continue$
                \ELSE
                    \STATE $\tau \leftarrow j$
                    \STATE $break$
                \ENDIF
            \ELSE
                \IF {$\psi_i, \varphi_j \neq -1$ and $a^1_i=\left\langle blk \right\rangle$ or $b^1_j=\left\langle blk \right\rangle$}
                    \STATE $i,j \leftarrow FindNext(\varphi_i, \psi_j, \Phi, \Psi)$
                    \STATE $continue$
                \ELSIF{$\psi_i= -1$, $a^1_i=\left\langle blk \right\rangle$, $a^2_i=b^1_j$ \\or $\varphi_j = -1$, $b^1_j=\left\langle blk \right\rangle$, $b^2_j=a^1_i$}
                    \STATE $i \leftarrow i+1$, $j \leftarrow j+1$
                    \STATE $continue$ 
                \ELSE
                    \STATE $\tau \leftarrow j$
                    \STATE $break$ 
                \ENDIF
            \ENDIF
        \ENDWHILE
        \STATE $\tau \leftarrow \mathcal{T}\ $ if\  $\tau=0$
        \RETURN $\tau$
    \end{algorithmic}
    \end{algorithm}
\vspace{-0.4cm} 

\section{Experiments and results}
\subsection{Experiment setup and dataset}

  We perform our experiments on a public English speech corpus, Librispeech, which contains 960 hours of audio in the training set. During fine-tuning, we only use 100 hours training subset and validate on dev-clean sets. Our results are reported on the test-clean and test-other sets. 
  
  All models are fine-tuned on the wav2vec2.0 pretrained model with CTC loss using the fairseq toolkit \cite{ott2019fairseq}. We use an Adam optimizer and a tri-stage rate scheduler with a phase ratio of 1:4:6. The peak of learning rate is set to 2e-5 and the max update number is 80000. We use the same settings as \cite{baevski2020wav2vec} for decoding, i.e., LM weight is 2.15 and word insertion penalty is -0.52. A 4-gram language model is used to do beam search and the beam size is fixed to 500. %Original wav2vec2.0 model has 7 convolution layers before the Transformer encoder, which downsamples the raw wave to 20ms per frame. 
  We change the convolution layer before the encoder blocks to a causal convolution layer with a kernel size of 15 for streaming ASR. The latency induced by the encoder (EIL) \cite{shi2021emformer} in our model is 0, so we use real-time-factor (RTF) to measure the increase in processing time caused by the revision strategy\footnote{Strictly speaking, RTF is often greater than 1.0 for streaming systems because of waiting for speech input. We ignore the speech input time and divide the average processing time by total audio length as RTF.}. The server contains a NVIDIA Tesla A30 24G GPU and Intel(R) Xeon(R) E5-2680 v4 @ 2.40GHz CPU.

\subsection{Baseline}

  Firstly, we try to reproduce the results in \cite{cao2021improving} as a baseline comparison, considering there are just a few studies on streaming models based on wav2vec2.0 fine-tuning. We keep the same settings as \cite{cao2021improving} and train a non-streaming model and a chunk-based streaming model with a chunk size of 0.96s (EIL is 0.48s) using the chunk mask given in Figure~\ref{fig:attention_mask}(b). We also train a causal model using the causal mask in Figure~\ref{fig:attention_mask}(a). The result is shown in Table~\ref{tab:revision} which indicates that the causal model performs much worse than the non-causal models.

\begin{table*}[t]
\caption{WERs and RTF of the baseline and different revision strategies with test-clean/test-other set. RTF is calculated based on the test-other set. 0.96-0.4 revision model refers to the model revision step is 0.96s and the interval is 0.4s.}
\label{tab:revision}
\centering
\begin{tabular}{ccccccc}
\hline
&\textbf{model} & \textbf{step} & \textbf{interval} & \textbf{test-clean} & \textbf{test-other} & \textbf{RTF} \\ 
\hline
& non-streaming model \cite{baevski2020wav2vec}  & 0s & /    & 3.4   & 8.0 & /\\
& non-streaming model  & 0s & / & 3.5                 & 8.8            & /    \\
\textit{Baseline}& knowledge distillation streaming model \cite{cao2021improving} & 0s & / & 3.7 & 9.8 &/ \\
& 0.96s chunk-based streaming model  & 0s & / & 4.0                 & 12.0       & /        \\
&causal model   & 0s   & /   & 5.5   & 16.8   & 0.0965  \\ 
\hline
&   & 1s  & 0s   & 3.9   & 10.6   & 1.4956  \\
&   & 1s  & 0.2s  & 3.9   & 10.9   & 0.2458  \\
&non-streaming model  & 1s  & 0.4s  & 4.0   & 11.2   & 0.1795  \\
\textit{Proposed}& +streaming decode  & 1s  & 0.6s  & 4.1   & 11.6   & 0.1581  \\
&   & 1s  & 0.8s  & 4.3   & 12.4   & 0.1521  \\
&   & 2s & 0.4s  & 3.7   & 9.8    & 0.2265  \\ 
\hline
&0.96-0.4 revision model  & 0.96s    & 0.4s                & 3.7               & 10.2   & 0.1682        \\ 
\textit{Proposed}&1-0.4 revision model  & 1s    & 0.4s                & 3.7               & 10.2  & 0.1714             \\
&2-0.4 revision model  & \textbf{2s}  & \textbf{0.4s}  & \textbf{3.7} & \textbf{9.2}       & 0.2247             \\
\hline
\end{tabular}
\end{table*}
% \vspace{-0.1cm}

\subsection{Encoder revision strategy}

  To choose a proper revision step and interval, we first use a non-streaming model and decode it in a streaming way to make trade-offs between the word error rates (WERs) and the time costs. We evaluate the influence of the size of history states \(h\) when applying the revision methods. When using the cached states and retaining the same revision settings, utilizing all history states has a relative increase of 8\% time costs in forward computation but has an 18\% relative WERs improvement over using 2s history states. So in subsequent experiments, we will save all historical states (\(h=t-1\) in Eq. \ref{eq2}) for inference. 
  
  The revision step is 1s to resemble the calculated length in baseline, and the revision interval is set from 0 to 0.8s in increments of 0.2s. 0 means the model will update previous states of 1s every frame. The result is shown in Table~\ref{tab:revision} which demonstrates that a smaller revision interval brings better performance, but at the same time, the overall time costs increase drastically according to Eq. \ref{eq16}. Considering RTF and WERs, the revision interval of 0.4s is an appropriate compromise for our method and subsequent experiments are based on this parameter setting. 
  
  To match the training and inference, we use a revision mask described in Figure~\ref{fig:attention_mask}(c) to train revision models. The revision interval is fixed to 0.4s, and we use 0.96s, 1s, 2s for the revision step. The result is shown in Table~\ref{tab:revision}. The performance with 2s revision step is 9.2 WERs on test-other set, which is much better because more revision steps provide more information for computation. Figure~\ref{fig:output} shows an example of decoding with 1s revision step and 0.4s revision interval, indicating that our method does not require an overall adjustment at the end of the sentence to achieve high accuracy, which guarantees stable prior outputs with low latency.
% \vspace{0.3cm}
\begin{table}[t]
\caption{The effect of different thresholds on WERs and RTF.}
\label{tab:time}
\centering
\begin{tabular}{ccccc}
\hline
\textbf{model}  & \textbf{$\boldsymbol{\theta}$} & \textbf{test-clean} & \textbf{test-other} & \textbf{RTF} \\
\hline
& 0.2   & 5.9           & 17.9            &0.0711 \\
& 0.3   & 5.7           & 17.3            &0.0718 \\
causal model &0.4   & 5.6           & 17.0            &0.0720 \\
+ PSD & 0.5   & 5.6           & 16.9            &0.0749 \\
&0.6   & 5.6           & 16.9            &0.0751 \\
& 0.7   & 5.5           & 16.8            &0.0800 \\
\hline
                & 0.2   & 3.9   & 10.9  & 0.1008       \\
                & 0.3   & 3.8   & 10.6  & 0.1104       \\
revision-only model               & 0.4   & 3.7            & 10.4           & 0.1224       \\
(1-0.4 revision model)              & 0.5   & 3.7            & 10.3           & 0.1327       \\
                & 0.6   & 3.7            & 10.3           & 0.1452       \\
                & 0.7   & 3.7            & 10.2           & 0.1528       \\
\hline
                & 0.2   & 4.0              & 11.4           & 0.1021       \\
                & 0.3   & 3.8   & 10.9  & 0.1060       \\
causal-revision model               & 0.4   & 3.8            & 10.7           & 0.1144       \\
(1-0.4 revision)              & 0.5   & 3.7            & 10.6           & 0.1228       \\
                & 0.6   & 3.7            & 10.6           & 0.1325       \\
                & 0.7   & 3.7            & 10.5           & 0.1406       \\
\hline
\end{tabular}
\end{table}
\vspace{-0.2cm}

\subsection{Accelerating revision strategy}

  Then we test the CTC spike position alignment decoding algorithm on 1-0.4 revision model in Table~\ref{tab:time}. The bigger the \(\theta\) sets, the stricter the selection of dominant label is. When \(\theta\) is 0.3, RTF greatly reduces (35.0\% relative) with a slight WERs increase (3.9\% relative), which is a proper compromise. We also decode on the causal model as a comparison since \(\theta\) will affect the usage of PSD. The results in Table~\ref{tab:time} show that applying PSD for causal model can reduce RTF while increasing WERs.
  
\begin{figure}[t]
\centering
\includegraphics[width=\linewidth]{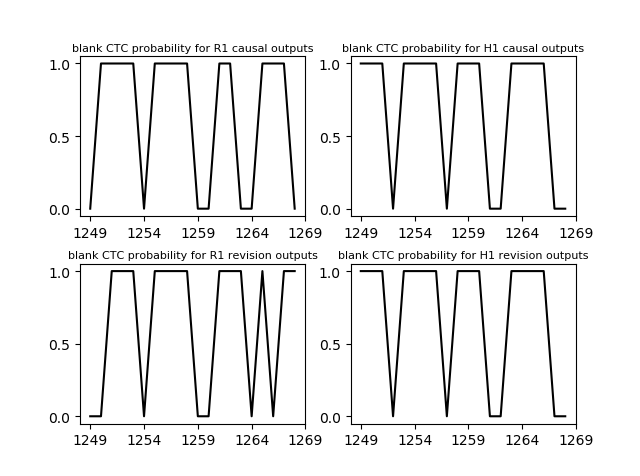}
\caption{Comparison between causal outputs and revision outputs based on revision-only model (R1) and causal-revision model (H1). Results are calculated by blank label dominant probabilities in 20 frames.}
\label{fig:CTCspike}
\end{figure}

  Causal outputs of the revision model contain a large number of \(\left\langle blk \right\rangle\) instead of actual labels due to training mismatches and are severely misaligned with the output CTC spikes of the revision part, which is not reasonable in practical applications. To solve this misalignment, we apply a causal-revision dynamic mask training inspired by \cite{zhang2020unified}, randomly choosing a causal mask or a revision mask in a proportion of 3:7 at each iteration. %We use causal and revision with a probability of 30\% and 70\%, respectively.  and train for 120,000 max updates to converge. 
  In Figure~\ref{fig:CTCspike} we plot the CTC blank probabilities calculated by dominant labels of causal outputs and revision outputs from revision-only and causal-revision models. As shown in Figure~\ref{fig:CTCspike}, there is a large misalignment in the causal and revision outputs for the revision-only model, while the spikes are basically aligned after switching to the causal-revision model. Under a premise of a little performance degradation, the RTF of the causal-revision model is lower than the revision-only model with the same threshold in Table~\ref{tab:time}, indicating that the causal and revision part output spikes are more aligned and fewer decoding paths are actually changed. 
  
  The surprisingly outstanding performance of revision-only model on RTF (especially when \(\theta\) is 0.2) is mostly determined by the PSD method that we also used in causal outputs. This causes frames to be skipped when outputting a lot of \(\left\langle blk \right\rangle\) if the causal part underperforms. We test the RTF when \(\theta\) is 0.2 both on revision-only and causal-revision models without PSD in causal outputs. RTF increases 33\% relatively for revision-only model and only 16\% relatively for causal-revision model which confirms our conjecture.

\section{Conclusion}
  In this paper, we apply an encoder states revision strategy to improve the causal models. To reduce the time costs caused by revision, we propose a CTC spike position alignment decoding algorithm. The results can best achieve 3.7/9.2 WERs, demonstrating that our method can be competitive with other non-causal models at the speed of causal models. Our future work will investigate how to align the CTC spike positions between causal and revision outputs without any accuracy loss.%The results achieve 3.8/10.9 WERs on test-clean/other sets considering the time costs and performance trade-off and can best achieve 3.7/9.2 WERs which demonstrate that our method can achieve comparable performance with chunk-based models and distillation models. Our future work will investigate how to align the CTC spike positions between causal outputs and revision outputs without any accuracy loss.

\bibliographystyle{IEEEtran}

\bibliography{mybib}

% \begin{thebibliography}{9}
% \bibitem[1]{Davis80-COP}
%   S.\ B.\ Davis and P.\ Mermelstein,
%   ``Comparison of parametric representation for monosyllabic word recognition in continuously spoken sentences,''
%   \textit{IEEE Transactions on Acoustics, Speech and Signal Processing}, vol.~28, no.~4, pp.~357--366, 1980.
% \bibitem[2]{Rabiner89-ATO}
%   L.\ R.\ Rabiner,
%   ``A tutorial on hidden Markov models and selected applications in speech recognition,''
%   \textit{Proceedings of the IEEE}, vol.~77, no.~2, pp.~257-286, 1989.
% \bibitem[3]{Hastie09-TEO}
%   T.\ Hastie, R.\ Tibshirani, and J.\ Friedman,
%   \textit{The Elements of Statistical Learning -- Data Mining, Inference, and Prediction}.
%   New York: Springer, 2009.
% \bibitem[4]{YourName17-XXX}
%   F.\ Lastname1, F.\ Lastname2, and F.\ Lastname3,
%   ``Title of your INTERSPEECH 2022 publication,''
%   in \textit{Interspeech 2022 -- 23\textsuperscript{rd} Annual Conference of the International Speech Communication Association, September 18-22, Incheon, Korea, Proceedings, Proceedings}, 2022, pp.~100--104.
% \end{thebibliography}

\end{document}